\pdfoutput=1
\documentclass[twocolumn]{aastex6}

\begin{document}

\title{On the radial metallicity gradient and radial migration effect of the Galactic disk}



\author{Yunpeng Jia\altaffilmark{1,2}, Yuqin Chen\altaffilmark{1,2},Gang Zhao\altaffilmark{1,2},
Xiangxiang Xue\altaffilmark{1},Jingkun Zhao\altaffilmark{1},
Chengqun Yang\altaffilmark{1,2}, Chengdong Li\altaffilmark{1,2}}

\altaffiltext{1}{Key Laboratory of Optical Astronomy, National Astronomical Observatories,
Chinese Academy of Sciences, Beijing, 100101, China}
\altaffiltext{2}{School of Astronomy and Space Science, University of Chinese Academy of Sciences, Beijing 100049, China}

\begin{abstract}
We study the radial metallicity gradient $\Delta[M/H]/\Delta R_g$
as a function of [Mg/Fe] and $|Z|$  with the help of a guiding radius based on 
the Apache Point Observatory Galactic Evolution Experiment and Gaia
 and then analyze the radial migration effect on
the radial metallicity gradient and metallicity-rotation gradient
between the Galactic thin and thick disks.
 The derived trend of gradient $\Delta[M/H]/\Delta R_g$ versus [Mg/Fe] shows a transition at
[Mg/Fe] $\sim 0.18$ dex, below which the gradient is
 negative and varies a little as [Mg/Fe] increases; however, it
 changes sharply in [Mg/Fe] ranges of 0.16-0.18, above which the gradient increases linearly with
 increasing [Mg/Fe], being a positive value at [Mg/Fe]$\gtrsim 0.22$ dex.
These positive gradients in the high-[Mg/Fe] populations are found at $|Z| < 0.8$ kpc,
and there are nearly no gradients toward higher $|Z|$.
By comparing the metallicity distributions, the radial metallicity gradients $\Delta[M/H]/\Delta R$
and the metallicity-rotation gradients between the total sample and $|R-R_g|<2$ kpc subsample
(or $|R-R_g|>2$ kpc subsample), we find that, for the thick disk, 
blurring flattens the gradient $\Delta[M/H]/\Delta R$
and favors metal-poor high-eccentricity stars.
These stars are responsible for the measured positive metallicity-rotation gradient
of the thick disk. 

\end{abstract}


\section{INTRODUCTION}
The Milky Way disk has been suggested to host a thick disk, in addition to its thin disk,
based on tje results of geometric decomposition works
\citep[e.g.][]{Gilmore-Reid1983,Juric2008,Jia2014}.
Generally, the thick disk is thought to be old as the thin disk \citep[e.g.][]{bensby05,yoachim08}.
Other age-related properties, such as $\alpha$-element abundances
 \citep[e.g.][]{fuhrmann1998, bensby03,bensby05, reddy06, haywood08,Adibekyan2011} and
 kinematics \citep{Chiba2000,Gilmore2002,Yoachim05},
are also found to be different from the thin disk.
 Although the formation mechanism of the thick disk is still debated,
four kinds of scenarios are widely known so far:
heating \citep[e.g.][]{Quinn1993}, accretion \citep[e.g.][]{Abadi2003},
gas-rich mergers \citep[e.g.][]{Brook2004,Bournaud2009}, and
radial migration \citep[e.g.][]{Schonrich-Binney2009}.
In particular, the last one has attracted much attention
because it could be triggered by the well-known
non-axisymmetric structures of the Milky Way disk (such as spiral arms and bars).
On the other hand, radial migration would alter the disk structure and chemical composition
through churning (stars changing angular momentum)
and blurring \citep[stars conserving their angular momentum,][]{Schonrich-Binney2009},
which makes the interpretation of the observational results complicated.
In this respect, the study on the radial metallicity gradient, which
records the enrichment history of stellar abundances at a given radial location,
provides a good way to disentangle the radial migration effect on the chemical evolution
of the Galactic disk.

The radial metallicity gradient has been well studied in previous works
based on a variety of stellar tracers. \citet{Tas2016} gave
a detailed summary in their Table 1, where
the thin disk has a negative gradient and the thick disk
shows a flat or even positive gradient. Moreover,
the overall disk gradient was found to gradually flattens toward high $|Z|$, which is proposed to
be driven mainly by radial migration \citep{Schonrich2017}.
Interestingly, the metallicity-rotation gradient is also found to be negative
for the thin disk but positive for the thick disk
\citep[e.g.][]{Adibekyan2013, Allende-Prieto2016}, which is also linked with the
radial metallicity gradient. It has been suggested that
the negative (positive) metallicity-rotation gradients
is driven by the negative (positive) radial metallicity gradient for thin (thick) disk stars
due to blurring effect \citep{Vera-Ciro2014,Allende-Prieto2016,Schonrich2017}.
In this case, a thick disk star born in the inner disk with epicyclic motion
would tend to be metal poor due to the positive radial metallicity gradient.
Meanwhile, its current location in the solar neighborhood is near
the orbit's apo-center and thus tends to rotate slower.
This naturally results in a positive metallicity-rotation gradient for thick disk stars.
Supporting this scenario, \citet{Toyouchi2014} reported that the measured radial metallicity gradients
for $\alpha$-enhanced (thick disk) populations are positive.

Note that \citet{Toyouchi2014} adopted the guiding radius $R_g$,
instead of current Galactocentric distance $R$, since
the gradient, if measured with $R$, likely suffers
from the radial migration effect and the usage of the guiding radius $R_g$
can diminish the blurring effect. However, the elemental abundances in \citet{Toyouchi2014}
are based on low-resolution spectra of the Sloan Digital Sky Survey (SDSS), and distances are based
on photometric methods.  These can be greatly improved using recently released high-quality data.
Investigating the radial metallicity gradient by taking into
account the guiding radius and the metallicity-rotation gradients based on accurate
elemental abundances and radial velocity from high-resolution spectra provided by
the Apache Point Observatory Galactic Evolution Experiment \citep[APOGEE,][]{Majewski2017} survey
is of much interest. Additionally, the wealth astrometric information from Gaia \citep{Gaia2016}
enables us to perform a statistical analysis on the radial metallicity gradient with the
aid of the guiding radius.
Specifically, we aim to derive the radial metallicity gradient
as  a function of  [Mg/Fe] and $|Z|$ and
analyze the effect of radial migration on the radial metallicity gradient
and metallicity-rotation gradient between the Galactic thin and thick disks.

\section{Data}
APOGEE is a high-resolution ($R\sim 22,500$), near-infrared H-band (1.5-1.7 \micron)
spectroscopic survey targeting on red giants
selected with a cut of $(J-K_s)_0 \ge 0.5$ \citep{Zasowski2013}.
It provides precise radial velocities, stellar parameters,
and elemental abundances through self-designed pipelines \citep{Nidever2015,Holtzman2015, Garcia2016}.
Recently, the second Gaia data release \citep[DR2,][]{Gaia2018} provides parallaxes (distances)
and proper motions for billions of stars,
which contains most stars in the APOGEE/SDSS DR14 survey \citep{Abolfathi2017}.
Our sample of stars is selected by cross-matching between APOGEE DR14  and Gaia DR2.

\begin{figure}
\figurenum{1}
\plotone{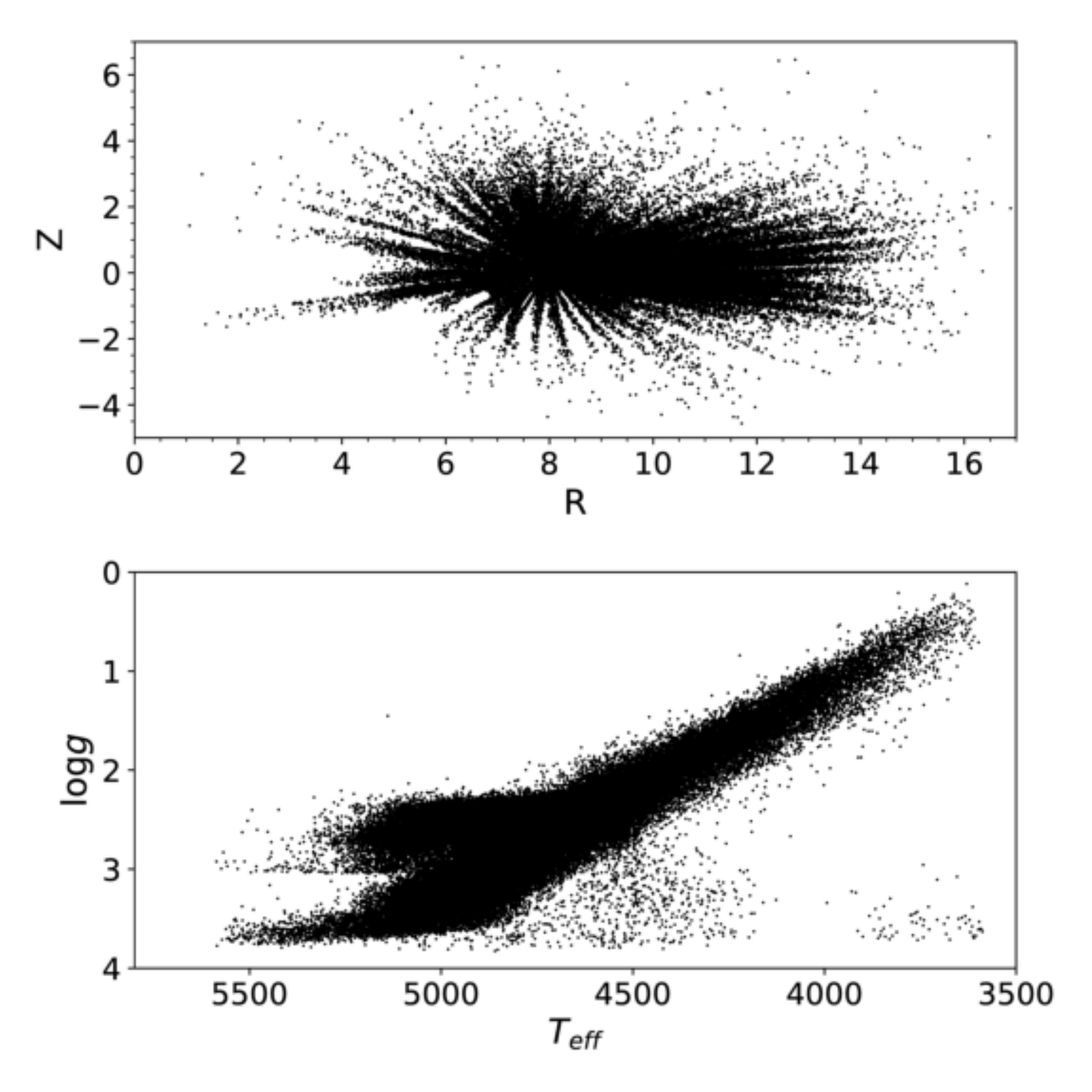}
\caption{Top panel: the spatial distribution of the sample stars in the Galactic cylindrical coordinates.
Bottom panel: the $T_{\rm eff}$ versus log$\,g$ diagram for the sample stars.}
\label{RZ}
\end{figure}

The distances in Galactic cylindrical coordinates $R$ and $Z$
are calculated using parallaxes from Gaia DR2, and we adopt the solar position of 8.0 kpc \citep{Reid1993}.
The spatial velocities ($U$,$V$,$W$) in left-hand Cartesian coordinates
are calculated from radial velocities from APOGEE DR14, with the distances and proper motions
from Gaia DR2. In the calculation, we restrict stars with the relative error of parallax less than
 20\%, which leads to the error of proper motion less than 1 $\rm mas\; yr^{-1}$ for most stars.
We correct the solar motion with respect to the local standard of rest
of $(U,V,W)_\sun$ = (-10.00, +5.25, +7.17) \citep{Dehnen1998}
and adopt coordinate transformation to obtain the cylindrical coordinate velocities
$V_R$ and $V_{\phi}$. The orbital parameters, peri-center ($R_p$), apo-center ($R_a$), and
eccentricity ($e=(R_a-R_p)/(R_a+R_p)$) are calculated with 
the \emph{galpy} \citep{Bovy2015} under the potential of \emph{MWPotential2014}.
Then, the guiding radius ($R_g$), at which the minimum of effective potential
occurs, is calculated under this potential through the following equation:
\begin{equation}
\frac{V^2_{circ}}{R_g} = \frac{L^2_z}{R^3_g},
\end{equation}
where $V_{\rm circ}$ is the circular velocity in this potential and
$L_z$ is the angular momentum, which is a constant of motion in axis symmetric potential.
The guiding radius identifies the radius of the circular orbit with an angular momentum of $L_z$.
 From the probability point of view, this radius better represents
the Galactocentric distance at which a star was born than $R$
since many stars do not have circular orbits \citep{Boeche2013}.
We then expect that a star with $R-R_g>0$ ($R-R_g<0$)
has moved outward (inward) to the current location after it was born.
In addition, we exclude stars with [M/H] of less than $-1.0$ dex and stars with a
total velocity of more than 150 $\rm kms^{-1}$ to diminish the contamination from the halo.
We also exclude stars at the fields of ($l$, $|b|$) $<$ (10$^\circ$, 10$^\circ$)
to remove bulge stars.
Finally, we have $147,794$ stars for the following analysis.
 The $R$ versus $|Z|$ plane and the $T_{\rm eff}$ versus log$\,g$ diagram of the sample
are shown in Fig.~\ref{RZ}.
This sample comprises giants with effective temperature in the range of 3500-5600 K, and
it covers a wide spatial range of $2<R<15$ kpc and $0<|Z|<5$ kpc.

\section{Analysis and Results}

\subsection{Radial metallicity gradient as a function of [Mg/Fe]}

\begin{figure*}
\figurenum{2}
\plotone{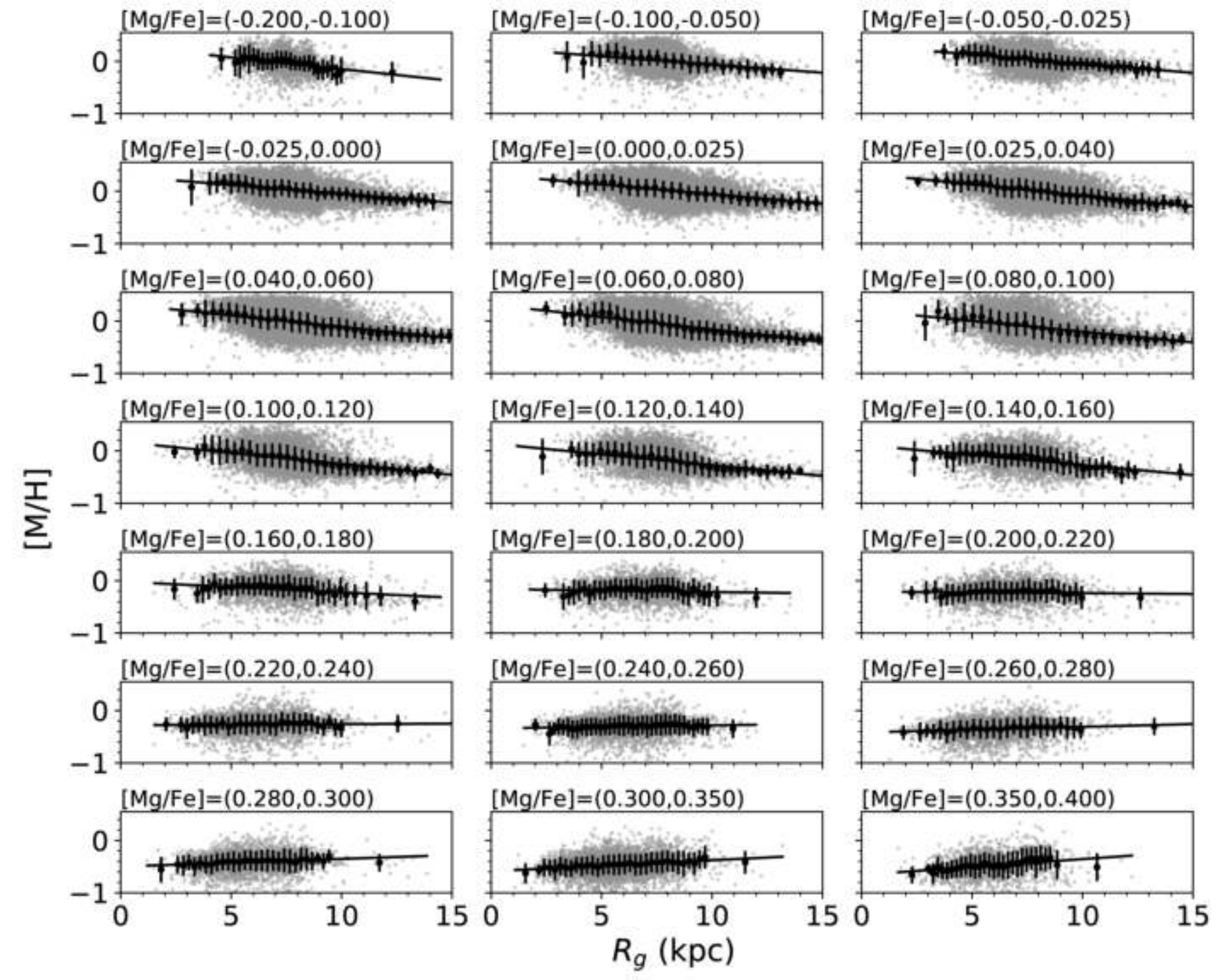}
\caption{$R_g$ versus [M/H] diagrams in 21 [Mg/Fe] bins.
The bin size is arbitrarily set to include enough stars used for statistical analysis.
In every bin, a linear regression is adopted to derive the radial metallicity gradient.
The black dots indicate the means of metallicities,
and the corresponding error bars represent the standard deviations of the means.}
\label{RgMh}
\end{figure*}

\begin{figure}
\figurenum{3}
\plotone{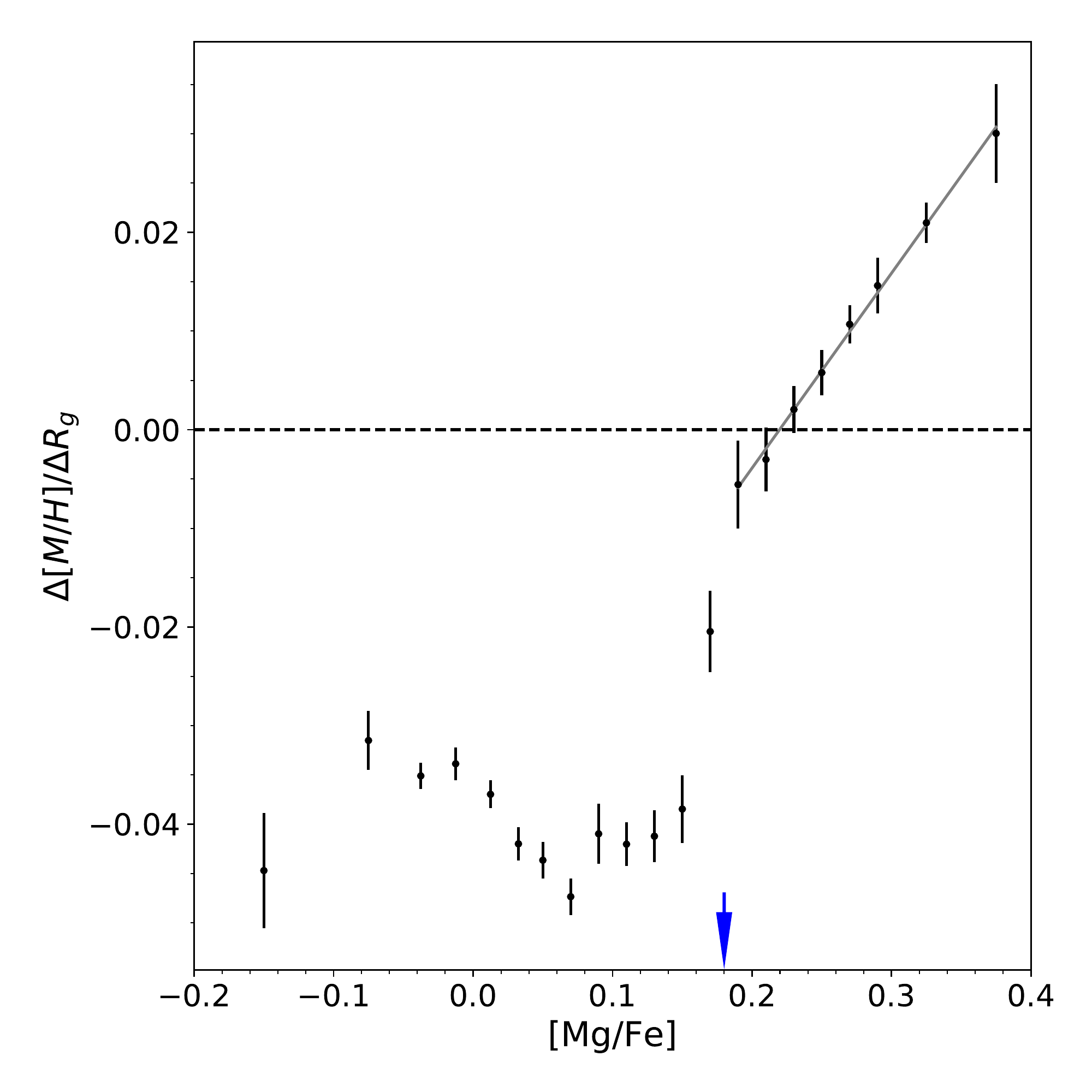}
\caption{Variation of the radial metallicity gradient with the [Mg/Fe] ratio.
 The gray straight line shows the rising trend of the high-[Mg/Fe] sequence, and
 the blue arrow indicates the transition between the thin and thick disk at [Mg/Fe] $\sim 0.18$ dex. }
\label{grad}
\end{figure}

In order to investigate the variation of metallicity gradient with [Mg/Fe],
we illustrate the $R_g$ versus [M/H] diagrams for 21 [Mg/Fe]
bins in Fig.~\ref{RgMh}. The bin size is arbitrarily set so that enough
stars can be used for the statistical analysis. 
In every [Mg/Fe] bin, we first divide $R_g$ into
50 equal-length intervals and then adjust adjacent intervals
to ensure the number of stars in every final interval is larger than 20.
In Fig.~\ref{RgMh}, black dots indicate the means of metallicities,
and the corresponding error bars represent the standard deviations of the means.
Then, we adopt a linear regression to the dots in every [Mg/Fe] bin and
obtain the radial metallicity gradient.
The resulting gradient varies from $-0.045$ dex/kpc to $+0.030$ dex/kpc
as [Mg/Fe] increases from $-0.20$ to $+0.40$ dex,  which is illustrated in Fig.~\ref{grad}.
Fig.~\ref{grad} shows that the gradient for low-[Mg/Fe] population with [Mg/Fe]
below $0.16$ dex is negative and varies slightly as [Mg/Fe] increases.
Then, the gradient changes sharply at [Mg/Fe] around 0.18 dex,
after which the gradient for the high-[Mg/Fe] population rises steadily
and becomes positive at [Mg/Fe]$\ge 0.22$.
 The absolute values of Pearson correlation coefficients are larger than 0.67
 with the $p$ values less than 0.001 for bins of [Mg/Fe]$\le 0.18$ dex and
 bins of [Mg/Fe]$\ge 0.26$ dex, indicating strong evidence of
 correlations between metallicity and $R_g$.
 For the middle [Mg/Fe] bin of 0.24-0.26 dex,
the coefficient is of 0.39 with a $p$ value of 0.017, indicating evidence of
correlation between metallicity and $R_g$. However, for the [Mg/Fe] bins of 0.18-0.24 dex, the coefficients
 are in the range of 0.16-0.22 with the $p$ values larger than 0.1,
 indicating no significant evidence for such correlations.
Note that the blue arrow in Fig.~\ref{grad} indicates a transition
 of the radial metallicity gradient at [Mg/Fe]$\sim 0.18$. In light of
 the fact that [Mg/Fe] is a proxy for age, this transition implies
the separation between the thick disk and the thin disk because
the former is typically older than the latter.
In short, the thick disk that is represented by stars with [Mg/Fe]$\ge 0.18$
has a linear trend in the gradient versus the [Mg/Fe] plane, and the thin disk
characterized by [Mg/Fe]$<$0.18 appears to be relatively complicated.
This difference indicates that the thin disk and the thick disk
are distinct stellar populations.

Our data is consistent with the results summarized in Table 1 of \citet{Tas2016}
in which the thin disk has a negative radial metallicity gradient,
whereas the thick disk has a flat or positive gradient.
 But the measured values of radial metallicity gradients for thick disk stars
are very different in the literature,
varying from a positive gradient \citep{Casagrande2011,Carrell2012,Tas2016},
to nearly no gradient
\citep{Cheng2012,Boeche2013,Boeche2014,Mikolaitis2014,Xiang2015,Plevne2015,Anders2017},
and even a negative gradient \citep{Hayden2014}.
These discrepancies may be caused by the sample biases.
As pointed out in \citet{Anders2017}, the  gradient could depend on,
e.g. the sample's spatial location, age distributions, insufficient statistics, and selection biases.
These factors are hardly to qualify among the works, and thus it is not easy to compare
our results with other observational works based on a different data sample and different tracers.
Instead, we prefer to compare our results with theoretical works.
In this respect, \citet{Kawata2018} performed $N$-body simulations to the APOGEE data and
found that the initial radial metallicity gradient
of the thick disk progenitor should not be negative, but should be either flat or even positive.
Moreover, some Galactic chemical evolution models \citep[e.g.][]{Chiappini2001,Schonrich2017}
suggest that the inside-out formation of the Galaxy
gives a positive radial gradient for thick disk stars.
Our results support these theoretical works by providing evidence that the radial
metallicity gradient of the Galactic disk depends on the [Mg/Fe] ratio, and the thick disk has
a flat or positive radial metallicity gradient.

\subsection{The radial metallicity gradients at different $|Z|$ intervals} \label{gradz}
\begin{figure}
\figurenum{4}
\plotone{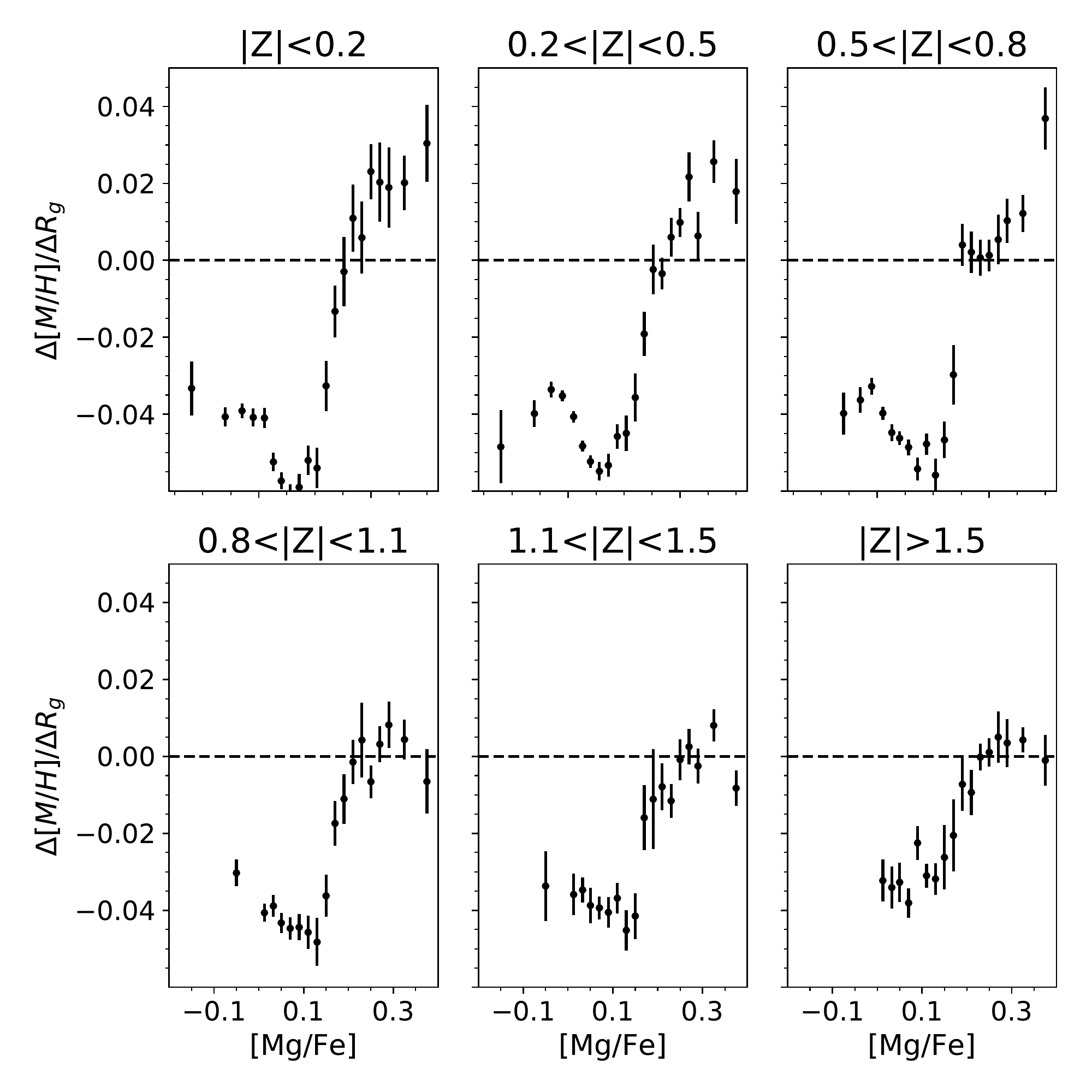}
\caption{Variation of radial metallicity gradients with [Mg/Fe] in different $|Z|$ bins.}
\label{grad-z}
\end{figure}

It is interesting to investigate the dependence of the relation between the radial metallicity gradient
and the [Mg/Fe] ratio on $|Z|$
in light of the fact that the thin and thick disks dominate at different $|Z|$ ranges.
For this purpose, we split $|Z|$ into six bins: $|Z|<0.2$ kpc, $0.2 < |Z| < 0.5$ kpc,
$0.5 < |Z| < 0.8$ kpc, $0.8 < |Z| < 1.1$ kpc, $1.1 < |Z| < 1.5$ kpc and $|Z| > 1.5$ kpc.
The relations of radial metallicity gradients versus [Mg/Fe] ratios at different $|Z|$ bins are
presented in Fig.~\ref{grad-z}.  
It shows that the negative gradient for
the low-[Mg/Fe] sequence remains in all $|Z|$ bins, while the
positive radial metallicity gradient of
the high-[Mg/Fe] sequence with [Mg/Fe]$\gtrsim 0.18$ dex only exists at $|Z| < 0.8$ kpc,
and there is nearly no gradient in larger $|Z|$ bins.
It should be noted that this trend disagrees with the result in \citet{Toyouchi2014}, who found that
the positive gradient is nearly unchanged as $|Z|$ increases (see their Fig.~4).

\subsection{Radial metallicity gradient between the thin and the thick disks}
\begin{figure}
\figurenum{5}
\plotone{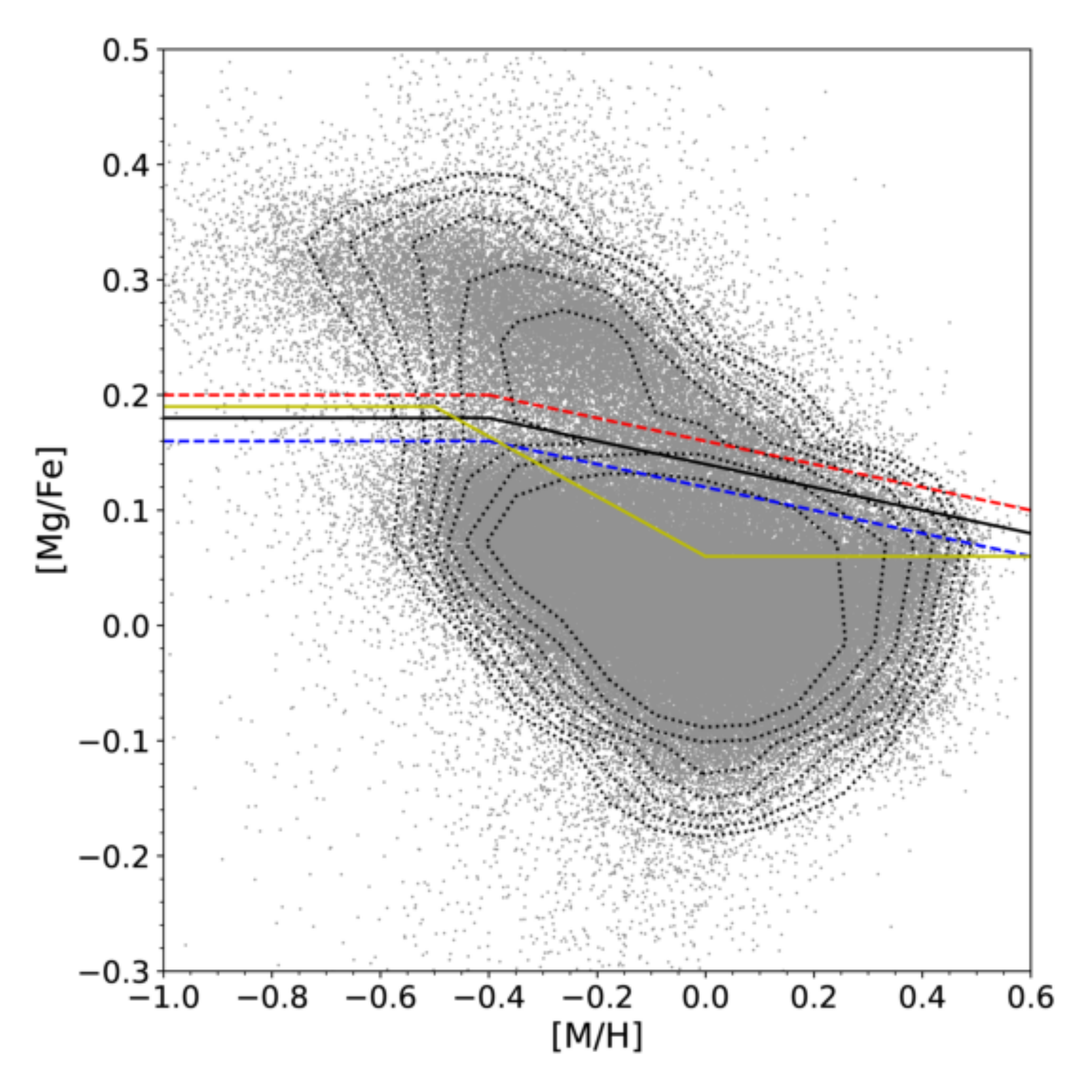}
\caption{[M/H] versus [Mg/Fe] diagram.
The black solid line cuts the sample into two star populations, i.e.
a thin disk population and a thick disk population.
This line is shifted by 0.02 dex in [Mg/Fe] to alter the contribution from the alternative disk.
The red line separates a relatively clean thick disk star population and the blue line
separates a relatively clean thin disk star population.
The yellow line marks the separation that used in the high-resolution study of \citet{Adibekyan2011}.
}
\label{mh-am-v1}
\end{figure}

It may be too simple to separate the thin disk and the thick disk based
on the [Mg/Fe] ratio at 0.18 dex only.  We adopt a conventional way to divide them
in the chemical plane, as shown in Fig.~\ref{mh-am-v1}.
A separation that adopted in the high-resolution study by \citet{Adibekyan2011} is
illustrated by the yellow line. It is clear that this separation is inappropriate for
this work.
Instead, we adopt a varying [Mg/Fe] ratio as a function
of [M/H] by passing the three points of (-1.0, 0.18), (-0.4, 0.18), and (0.6, 0.8)
in the [M/H] versus [Mg/Fe] diagram, which is presented by a black solid line in Fig.~\ref{mh-am-v1}.
We shift this line by 0.02 dex in [Mg/Fe] upward (red dash line) and downward (blue dash line)
in order to examine if the derived result
depends on the separation criterion between the thin and thick disk.
Note that the red line defines
a relatively clean thick disk star population, and the blue line
separates a relatively clean thin disk star population.

\begin{figure}
\figurenum{6}
\plotone{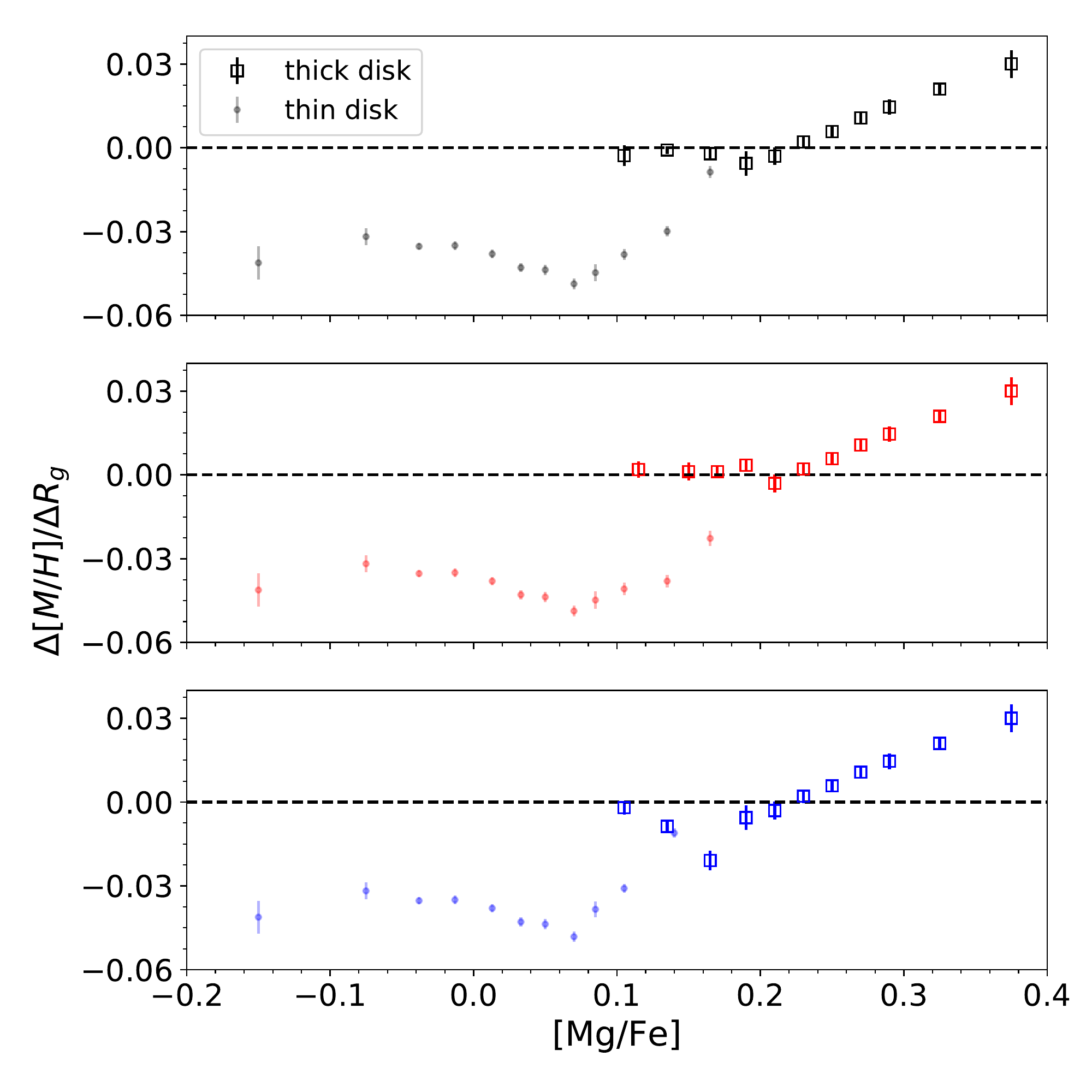}
\caption{Evolution of radial metallicity gradient with [Mg/Fe] for
thin and thick disks, which are marked by dots and squares, respectively.
Three panels correspond to the three solid lines in Fig.~\ref{mh-am-v1}
that used to divide the thin and thick disk populations.
The adopted colors are a one-to-one correspondence between this figure and Fig.~\ref{mh-am-v1}.
}
\label{grad-further}
\end{figure}

Based on the new separations, similar procedures are adopted to derived the
radial metallicity gradient as functions of
[Mg/Fe] for the thin disk and the thick disk populations in Fig.~\ref{grad-further}.
 From the top panel to the bottom panel, the results correspond to the separations that
represented by the black, red, and blue lines, respectively.
Compared with Fig.~\ref{grad}, a prominent feature of the top panel of this figure
 is that the thick disk presents a flat gradient at the low-[Mg/Fe] end
 and then subsequently rises linearly to the high-[Mg/Fe] end.
Similar features also hold in the middle panel that correspond to a separation
containing a relatively clean sample of thick disk stars.
 However, the bottom panel shows that
the flat gradients in the [Mg/Fe] range of 0.12-0.18 dex for the thick  disk
have been affected by pollution of thin disk stars and appear negative at [Mg/Fe] $\sim 0.17$ dex.
This suggests that adopting an improper chemical definition of the thick disk
 would affect the derived radial metallicity gradient.
Besides, we notice an interesting feature, which is that the gradient of the thin disk
  seems to transit continuously into the thick disk  by following the linearly rising trend.
This feature somewhat coincides with the finding that the structural parameters smoothly transit
in the chemical plane \citep{Bovy2012, Bovy2016}.

\subsection{Radial migration effect} \label{rme}
One superiority of using the guiding radius to derive radial metallicity gradient
is that it can diminish the blurring effect as it conserves the angular momentum.
We expect that the quantity of $R-R_g$ could somewhat reflect the radial migration effect,
and this quantity would be a simple and first approximation to help us disentangle how
the radial migration affect the observation results.
We assume that a subsample of stars with $|R-R_g|<2$ kpc would
suffer little from the radial migration effect (blurring), and 
the remaining subsample with $|R-R_g| >2$ kpc is supposed to include the migration stars.
In light of this, we expect that the footprint of the radial migration effect can be traced by comparing
the results derived from the $|R-R_g| <2$ kpc subsample (or $|R-R_g| >2$ kpc subsample) 
and the total sample. we note 
that the stars with $|R-R_g| >2$ kpc are tend to move on highly eccentric orbit.

\begin{figure}
\figurenum{7}
\plotone{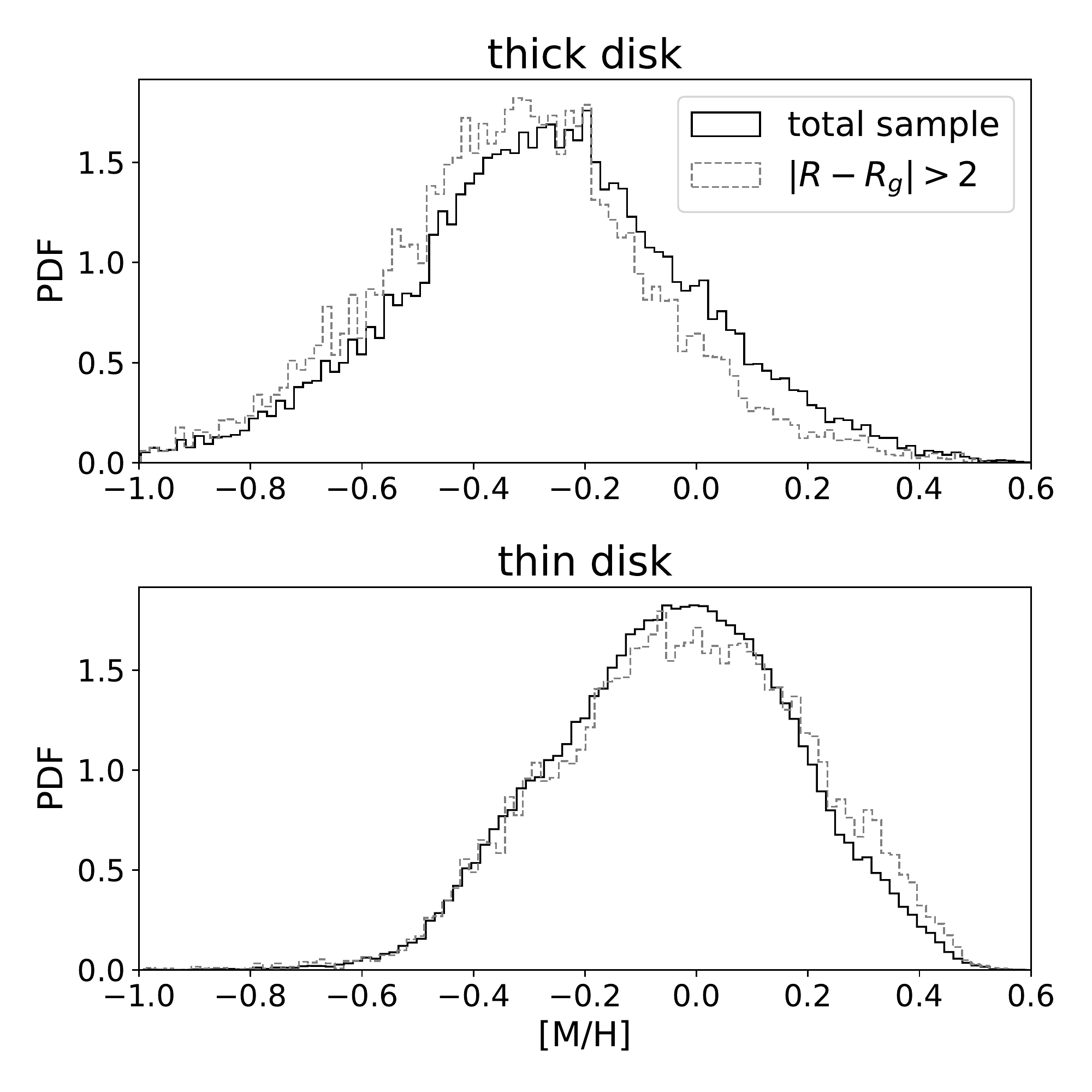}
\caption{Metallicity distributions of the total sample and the $|R-R_g| >2$ kpc subsample for
thick disk (top panel) and thin disk (bottom panel). }
\label{mh-dist}
\end{figure}

We first investigate the differences in the metallicity distributions
between the subsample of $|R-R_g| >2$ kpc and the total sample for the thin disk and the thick disk
in Fig.~\ref{mh-dist}. In the top panel of this figure,
metal-poor stars are over-populated in the $|R-R_g| >2$ kpc sample as compared to the total sample.
This implies that the migration favors the metal-poor for thick disk population, while
the bottom panel shows that the metallicity distributions are similar
between these two samples, which means that the migration does not play a significant
role in shaping the thin disk's metallicity distribution.

Second, we compare the radial metallicity gradients $\Delta[M/H]/\Delta R$ between
the $|R-R_g| <2$ kpc subsample and the total sample by
using Galactocentric distance $R$.
We find that the thin disk gives a similar negative gradient between these two samples.
However, the gradient $\Delta[M/H]/\Delta R$ in the thick disk becomes flat, changing
from $0.016\pm 0.003$ dex/kpc to $0.006\pm 0.003$ dex/kpc when
moving from the $|R-R_g|<2$ kpc subsample to the total sample.
Considering the total sample contains migration  stars with $|R-R_g| >2$ kpc,
this is an evidence that radial migration flattens the
radial metallicity gradient $\Delta[M/H]/\Delta R$.

\begin{figure}
\figurenum{8}
\plotone{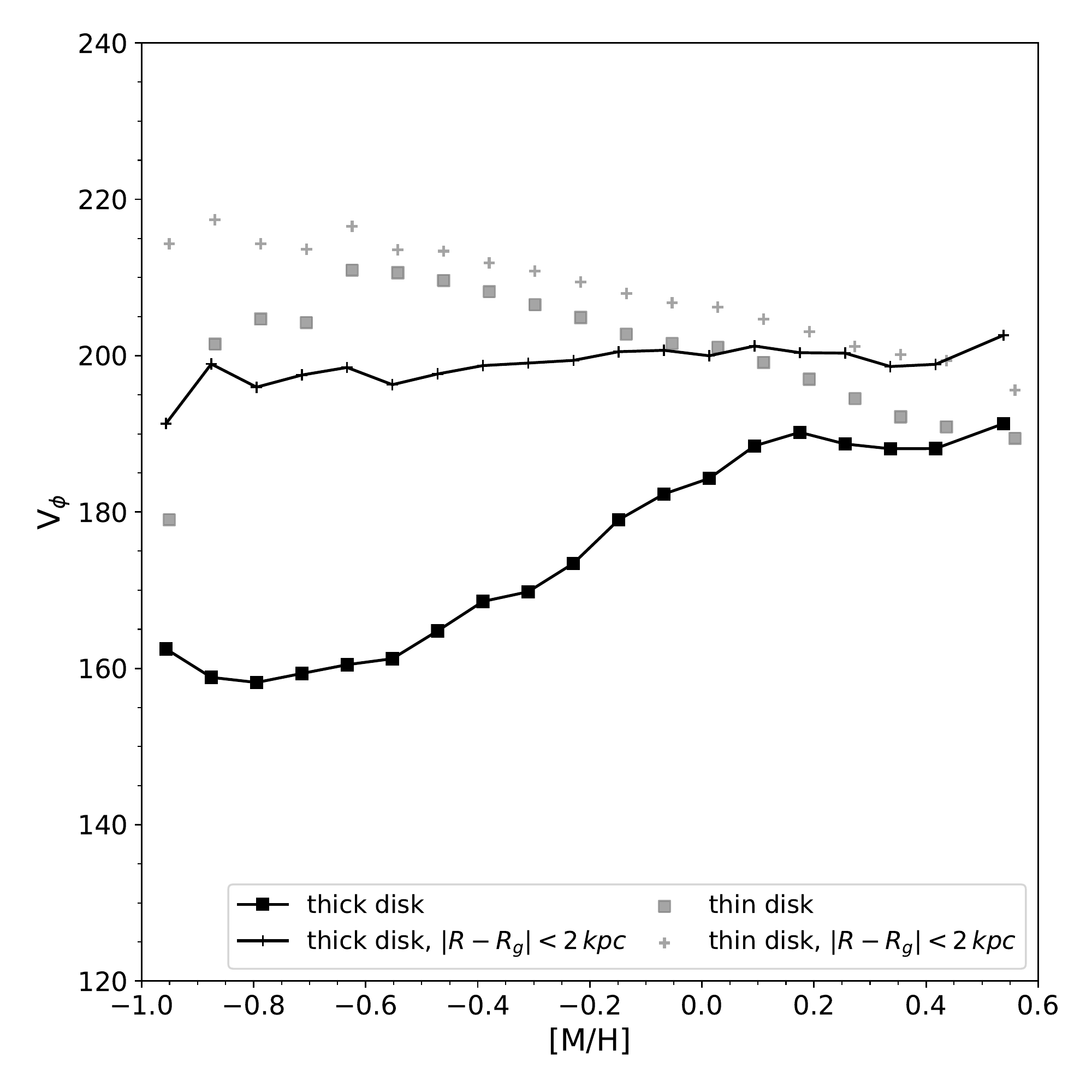}
\caption{Metallicity-rotation diagram for the total sample (marked by squares)
and the $|R-R_g| <2$ kpc subsample (marked by crosses). The thick disk stars
are highlighted with black lines to better illustrate
the change in the gradient $\Delta V_{\phi}/\Delta [M/H]$.}
\label{mh-rotation}
\end{figure}

Finally, we investigate the differences in the metallicity-rotation gradients
between the $|R-R_g| <2$ kpc subsample and the total sample in Fig.~\ref{mh-rotation}.
The total sample gives a positive gradient of 26.5 $\rm km\,s^{-1}\,dex^{-1}$ and a negative gradient of
-6.4 $\rm km\,s^{-1}\,dex^{-1}$ for thick disk and thin disk, respectively.
However, the gradient changes from 26.5 $\rm km\,s^{-1}\,dex^{-1}$ to 3.8 $\rm km\,s^{-1}\,dex^{-1}$
when using $|R-R_g| <2$ kpc subsample for the thick disk, which is close to
the gradient of -6.4 $\rm km\,s^{-1}\,dex^{-1}$ for the thin disk.
Note that the only difference between these two samples is the absence of
some metal-poor stars with high eccentricity in the $|R-R_g|<2$ kpc subsample.
Therefore, the change in the metallicity-rotation gradient for the thick disk
implies that the metal-poor high-eccentricity stars are responsible for the thick disk's
 positive $\Delta V_{\phi}/\Delta [M/H]$.
This is an observational signature in which blurring contributes to the positive
metallicity-rotation gradient of the thick disk.

\section{Implications on the formation and evolution of the disk} \label{discussion}

Before we discuss the implications of our results on the formation and evolution of the disk,
we note that the usage of the guiding radius,
rather than the Galactocentric distance $R$, may introduce a bias against
metal-rich inner disk stars that move on nearly circular orbits, as proposed by \citet{Boeche2013}.
This bias arises only when the eccentricity of the star is not well-sampled
at small $R_g$. Due to the large sample size in our work provided by
APOGEE and Gaia, the eccentricities are well-sampled at $R_g\gtrsim 4$ kpc,
and even at smaller $R_g$, a fraction of low-eccentricity stars still exist.
In view of this, we expect this bias would not affect much on our measured radial metallicity gradients.
Moreover, in the small $R_g$ region, the potential missing stars are supposed to be, 
on average, metal-rich \citep{Boeche2013} and should have relatively 
low-[Mg/Fe] ratios, thus we consider this bias in particular has less effect on 
the thick disk (high-[Mg/Fe] star population).

The variations of the radial metallicity gradient with [Mg/Fe] and $|Z|$
provide a way to infer the star formation history of the disk.
As shown in Fig.~\ref{grad} and Fig.~\ref{grad-further},
the thin disk shows negative radial metallicity gradient,
whereas the thick disk presents flat and positive gradients.
The negative gradient of the thin disk is usually understood as a natural result of the inside-out
formation scenario \citep[e.g.][]{Chiappini2001}, i.e. the star formation rate is
higher in the inner disk than the outer disk.
The positive gradient of the thick disk can also be achieved in this scenario under reasonable assumptions,
which was recently suggested by \citet{Schonrich2017}. They showed an example of
the achieved positive radial metallicity gradient for star-forming gas 
by adopting high central gas-loss rates and
re-accretion of one quarter of the lost enriched material in
conjunction with the inside-out formation.

However, the results presented in Sect.~\ref{gradz} suggest that the positive radial metallicity gradients
of the thick disk only exist in $|Z|<0.8$ kpc, and
the gradients are found to be flattened at larger $|Z|$ intervals complicates this formation scenario. 
These features seems to be reconciled with the upside-down disk formation fashion \citep[e.g.][]{Bird2013},
i.e. stars first forming in a geometrically thick layer and then in thinner layers.
A recent work from \citet{Kawata2018} 
suggested that the thick disk forming in an upside-down fashion
could have a different initial radial metallicity gradient, even including a negative one, and
then the effect of the inside-out formation produced an overall flat or positive
gradient when the formation was completed.
Another work by \citet{Fragkoudi2017}  supports this argument.
They suggested that the inner disk star populations arose from an interstellar medium
that has a mostly flat radial metallicity gradient forming in an upside-down fashion.
These works support a picture where the Galactic disk formed in an inside-out and upside-down fashion,
in which stars first formed in the thick layer (or larger $|Z|$)  had flat radial
metallicity gradients, and the subsequently formed stars in the thin layer (smaller $|Z|$)
obtained their positive gradients through the inside-out formation fashion,
thus the gradients are positive at low $|Z|$ and flat at larger $|Z|$.
Therefore, our results can be understood from the inside-out and upside-down
disk formation scenario.

\section{Summary and Conclusions}
In this work, we investigate the
radial metallicity gradient between the thin and thick disks, with the help
of the guiding radius, and analyze the effect of radial migration on the
derived radial metallicity gradient and the metallicity-rotation gradient based on
a large and high precision data from Gaia and APOGEE.
To infer the Galactic disk formation history, we derive the radial 
metallicity gradient $\Delta[M/H]/\Delta R_g$ as a function of [Mg/Fe] and $|Z|$.
We find that the radial metallicity gradient of low-[Mg/Fe] population ([Mg/Fe]$<$0.16 dex)
 is negative and has small variations with [Mg/Fe], and this
 gradient changes sharply until [Mg/Fe] reaches 0.18 dex,
 after which the gradient rises linearly and becomes positive at [Mg/Fe]$\ge 0.22$.
The positive gradients in the high-[Mg/Fe] population are found to limit at
$|Z|<0.8$ kpc only, and there is nearly no gradient toward higher $|Z|$.
The above results imply that the Galactic disk formed in an inside-out and upside-down
fashion. Moreover, the different trends of $\Delta[M/H]/\Delta R_g$ versus [Mg/Fe]
between the thick disk and thin disk, which are
characterized by the high-[Mg/Fe] population ([Mg/Fe]$>$0.18 dex) 
and the low-[Mg/Fe] population ([Mg/Fe]$<$0.18 dex), respectively,
indicate that the transition between them occurs at [Mg/Fe]$\sim$0.18 dex.
With a more specific division of the thin and thick disks in the chemical plane, 
we still find that the trends of $\Delta[M/H]/\Delta R_g$ versus [Mg/Fe]
are rather different between the thin and thick disks, which indicates that the
thin disk  and thick disk are two distinct star populations.
In the meantime, the thick disk clearly presented a
 flat or positive gradients depending on the [Mg/Fe] ratio.

In order to study the radial migration effect, we adopt a quantity, $|R-R_g|$,
to reflect the blurring effect since blurring does not change on $R_g$ but does on $R$. 
A subsample with $|R-R_g|<2$ kpc is thought
to not suffer from the blurring effect, and the remaining subsample with $|R-R_g|>2$ kpc
is considered to contain migration stars.
By comparing the metallicity distributions, the radial metallicity gradients $\Delta[M/H]/\Delta R$,
and the metallicity-rotation gradients between the total sample and
the alternative two above subsamples, we find that blurring process
flattens the gradient $\Delta[M/H]/\Delta R$
and favors metal-poor high-eccentricity stars for the thick disk.
These stars are responsible for the measured positive metallicity-rotation gradient
in the thick disk.

\section*{ACKNOWLEDGMENTS}
We would like to thank the referee for their constructive comments that
significantly improve this paper. 
This work is supported by the National Natural Science Foundation of China under grant
No. 11625313, 11573035, and 11390371; the Astronomical
Big Data Joint Research Center; and co-founded
by the National Astronomical Observatories, Chinese Academy of Sciences, and the Alibaba Cloud.
X.-X Xue thanks the "Recruitment Program of Global Youth Experts" of China.

Funding for the Sloan Digital Sky Survey IV has been provided by the Alfred P. Sloan Foundation, the U.S. Department of Energy Office of Science, and the Participating Institutions. SDSS-IV acknowledges
support and resources from the Center for High-Performance Computing at
the University of Utah. The SDSS web site is www.sdss.org.

SDSS-IV is managed by the Astrophysical Research Consortium for the
Participating Institutions of the SDSS Collaboration including the
Brazilian Participation Group, the Carnegie Institution for Science,
Carnegie Mellon University, the Chilean Participation Group, the French Participation Group, Harvard-Smithsonian Center for Astrophysics,
Instituto de Astrof\'isica de Canarias, The Johns Hopkins University,
Kavli Institute for the Physics and Mathematics of the Universe (IPMU) /
University of Tokyo, Lawrence Berkeley National Laboratory,
Leibniz Institut f\"ur Astrophysik Potsdam (AIP),
Max-Planck-Institut f\"ur Astronomie (MPIA Heidelberg),
Max-Planck-Institut f\"ur Astrophysik (MPA Garching),
Max-Planck-Institut f\"ur Extraterrestrische Physik (MPE),
National Astronomical Observatories of China, New Mexico State University,
New York University, University of Notre Dame,
Observat\'ario Nacional / MCTI, The Ohio State University,
Pennsylvania State University, Shanghai Astronomical Observatory,
United Kingdom Participation Group,
Universidad Nacional Aut\'onoma de M\'exico, University of Arizona,
University of Colorado Boulder, University of Oxford, University of Portsmouth,
University of Utah, University of Virginia, University of Washington, University of Wisconsin,
Vanderbilt University, and Yale University.

This work has made use of data from the European Space Agency (ESA) mission
{\it Gaia} (\url{https://www.cosmos.esa.int/gaia}), processed by the {\it Gaia}
Data Processing and Analysis Consortium (DPAC,
\url{https://www.cosmos.esa.int/web/gaia/dpac/consortium}). Funding for the DPAC
has been provided by national institutions, in particular the institutions
participating in the {\it Gaia} Multilateral Agreement.

\bibliographystyle{aasjournal}
\bibliography{mybibtex}

\end{document}